\documentstyle[11pt,a4]{article}

\newcommand{\be}{\begin{equation}}
\newcommand{\ee}{\end{equation}} 
\newcommand{\bea}{\begin{eqnarray}}
\newcommand{\eea}{\end{eqnarray}}

\begin{document}

\begin{titlepage}

\begin{flushright} 
{\tt FTUV/98-55\\ 
     IFIC/98-56 \\ 
     hep-th/9807019}
 \end{flushright}

\bigskip

\begin{center}

{\bf \LARGE A Note on Einstein Gravity on AdS$_3$ and \\ 
	Boundary Conformal Field Theory}
\footnote{Work partially supported by the 
{\it Comisi\'on Interministerial de Ciencia y Tecnolog\'{\i}a}\/ 
and {\it DGICYT}.}

\bigskip 

 J.~Navarro-Salas\footnote{\sc jnavarro@lie.uv.es} and
 P.~Navarro\footnote{\sc pnavarro@lie.uv.es}.

\end{center}

\bigskip

\begin{center}
\footnotesize
        Departamento de F\'{\i}sica Te\'orica and 
	IFIC, Centro Mixto Universidad de Valencia-CSIC.
	Facultad de F\'{\i}sica, Universidad de Valencia,	
        Burjassot-46100, Valencia, Spain. 
\end{center}               

\normalsize

\bigskip%\vskip2mm 
%\centerline{\today}
\bigskip%\vskip2mm

\begin{center}
			{\bf Abstract}
\end{center}
We find a simple relation between the first subleading terms
in the asymptotic expansion of the metric field in AdS$_3$, 
obeying the Brown-Henneaux boundary conditions, and
the stress tensor of the underlying Liouville theory on the
boundary. We can also provide an more explicit relation between the
bulk metric and the boundary conformal field theory when it
is described in terms of a free field with a background charge. 

\vspace{4cm}

\noindent PACS number(s): 04.60.Kz, 04.60.Ds

\end{titlepage}

\newpage

A crucial property of 2+1 dimensional gravity with a negative cosmological constant 
$\Lambda = -{1 \over \ell^2}$, described by the action
\be
  S= {1\over 16 \pi G} \int d^3 x \sqrt{-g} (R + {2\over \ell^2}) \, ,\label{action}
\ee
is that the asymptotic symmetry of the theory is the conformal algebra with a central charge given
by \cite{BH}
\be
   c={3\ell \over 2G} \, . \label{central}
\ee
Recently Strominger \cite{Strominger} has used this result to derive the Bekenstein-Hawking area 
formula for the 3d black holes \cite{BTZ} by counting the number of states of the conformal field
theory at infinity with central charge \label{c} via Cardy's formula \cite{Cardy}. Other approaches
\cite{Carlip} describe excitations associated to a conformal field theory at the horizon or at any
value of the black hole radius \cite{BBO}.

Taking into account that (\ref{action}) can be reformulated \cite{AT,Witten} as a Chern-Simons gauge theory
with gauge group $SL(2,R)\bigotimes SL(2,R)$, and using WZW reduction, one can argue that the asymptotic 
dynamics of (\ref{action}) is described by Liouville theory \cite{CHvD}. The aim of this letter is to further
elucidate the relation between the gravity theory and the underlying conformal field theory.

Following \cite{BH,Strominger} we assume the following asymptotic behaviour of the metric
\bea  
  g_{+-}& = & -{r^2\over 2} + {\gamma_{+-}(x^+,x^-)} + {\cal O} ({1 \over r}) \, , \label{g1} \\  
  g_{\pm\pm}& = & {\gamma_{\pm\pm}(x^+,x^-)} + {\cal O}({1 \over r}) \, , \label{g2} \\
g_{\pm r}& = & {{\gamma_{\pm r}(x^+,x^-)} \over r^3} + {\cal O} ({1 \over r^4}) \, , \label{g3} \\
g_{rr}& = & {\ell^2 \over r^2} + {{\gamma_{rr}(x^+,x^-)} \over r^4} + {\cal O} ({1 \over r^5}) \, ,
\label{g4}  
\eea
where $x^\pm \equiv {t \over \ell} \pm \theta$, and $\theta$ and $r$ are the angular and radial coordinates. 
We have introduced
explicitly the first subleading terms in the asymptotic expansion of the metric field and we want
to investigate how they are related to the conformal field theory on the boundary.
The infinitesimal diffeomorphisms  $\zeta^a (r, t,\theta)$preserving (3-6) are of the form 
  \cite{BH,Strominger}
\bea
\zeta^+ & = & 2T^+ + {\ell^2\over r^2} \partial_-^2 T^- + {\cal O} ({1\over r^4}) \, , \\ 
\zeta^- & = & 2T^- +{\ell^2 \over r^2} \partial_+^2 T^+ + {\cal O} ({1 \over r^4}) \, , \\
\zeta^r & = & -r(\partial_+ T^+ + \partial_- T^-) + {\cal O} ({1\over r}) \, ,
\eea
where the functions $T^\pm(x^\pm)$ verify the conditions $\partial_\pm T^\mp =0$.
Those diffeomorphisms with $T^\pm = 0$ should be considered as "gauge transformations". Therefore, if one 
consider the diffeomorphisms
\bea
\zeta^\pm & = & {{\alpha^\pm} \over r^4} + {\cal O} ({1 \over r^5}) \, , \label{gauge1} \\
\zeta^r & = & {{\alpha^r} \over r} + {\cal O} ({1 \over r^2}) \, , \label{gauge2}
\eea
where $\alpha^\pm$ and $\alpha^r$  are arbitrary functions of $x^+$ and $x^-$, it is not difficult to 
see that the variables $\gamma_{\pm\pm}$, $\gamma_{+-} - {1 \over 4\ell^2}\gamma_{rr}$
are the only gauge invariant quantities. Moreover, the equations of motion imply that
\be
0 = R + {6 \over \ell^2} = -{8\over r^2\ell^2} (\gamma_{+-} - {1 \over 4\ell^2}\gamma_{rr}) 
+ {\cal O}({1 \over r^3}) \, ,
\ee 
and this requires that
\be
\gamma_{+-} - {1 \over 4\ell^2}\gamma_{rr} = 0
\ee  
  The remaining equations of motion $R_{\mu\nu} - {1 \over 2} g_{\mu\nu} R = -{1 \over \ell^2} g_{\mu\nu}$                                       
lead to the equations
\bea
\partial_-\gamma_{++} = 0 \, , \\
\partial_+\gamma_{--} = 0 \, ,
\eea
  In addition one can make, using the gauge transformations (\ref{gauge1}) y (\ref{gauge2}), 
the following consistent gauge choice
\bea
\gamma_{\pm r} & = & 0 \, , \\
\gamma_{rr} & = & 0 \, ,
\eea
so the physical degrees of freedom are described by 
two chiral functions $\gamma_{\pm\pm}(x^\pm)$:
\be
ds^2 = {\ell^2 \over r^2} dr^2 - r^2 dx^+dx^- + \gamma_{++}(dx^+)^2 + \gamma_{--}(dx^-)^2 + 
{\cal O} ({1 \over r}) \, , \label{metric}
\ee
  For instance, the standard BTZ black hole solutions can be brought, via gauge transformations,
  to the form (\ref{metric}) with
\bea
\gamma_{++} & = & 2G\ell(M\ell+J) \, , \\
\gamma_{--} & = & 2G\ell(M\ell-J) \, ,
\eea 
  It is interesting to note that, when either $\gamma_{++} = 0$ or $\gamma_{--} = 0$,
the omitted terms in the expansion (\ref{metric}) vanish (in the appropriate gauge) and 
the corresponding BTZ solutions describe extremal black hole geometries. Hence
\be
ds^2 = {\ell^2 \over r^2} dr^2 - r^2 dx^+dx^- + \gamma_{++}(dx^+)^2 \label{chiral}
\ee
is an exact solution.

  We want now to identify the fields $\gamma_{\pm\pm}$ in terms of appropiate conformal fields 
on the boundary. To this end, let us determine the conformal transformation 
properties of these fields. The action of the diffeomorphisms (7-9) on the metric
(\ref{metric}) induces the following transformation law 
\be
\delta_{T^\pm}\gamma_{\pm\pm} = 2(T^\pm \partial_{\pm} \gamma_{\pm\pm} 
+ 2\gamma_{\pm\pm} \partial_{\pm} T^\pm) - \ell^2 \partial_{\pm}^3 T^{\pm} \, .
\label{delta}
\ee
  It is then clear that the variables $\gamma_{\pm\pm}$ are proportional, up to a constant ${-c\over 24}$, to the stress
tensor components $\Theta_{\pm\pm}$ of the underlying conformal field theory with 
central charge $c={3\ell \over 2G}$
\be
\Theta_{\pm\pm} = {1 \over 4\ell G}\gamma_{\pm\pm} + {\ell \over 16G} \, ,
\label{stresstensor}
\ee
  The above relation can also be confirmed by working out the conserved charges 
$J[\xi]$ given in \cite{BH}
\be
J[\xi] \propto \lim_{r \to \infty}\int d\phi\lbrace {\ell \over r}\xi^{\bot} +
{r^3 \over \ell^3}\xi^{\bot}(g_{rr} - {\ell^2 \over r^2}) +
{1 \over \ell}({\xi^{\bot} \over r} + \xi_{,r}^{\bot})(g_{\phi\phi} - r^2) +
2\xi^{\phi}\pi_{\phi}^r\rbrace \, ,
\ee 
in terms of the metric (\ref{metric}). One obtains
 \be
J[\xi] \propto \int d\phi \lbrace T^+(4\gamma_{++} + \ell^2) +
T^-(4\gamma_{--} + \ell^2) \rbrace \, ,
\ee
which corresponds to the conserved charges associated to the stress tensor 
(\ref{stresstensor}) of a conformal field theory on a cylinder with central 
charge $c={3\ell \over 2G}$. Note that the shift term ${c \over 24}$ in (\ref{stresstensor})
arises from changing variables from the sphere to the cylinder
$z=e^{{t \over \ell}+i\theta}$.
  The Fourier components $L_n (\overline{L}_n)$ of 
$\Theta_{++} (\Theta_{--})$ obey the Virasoro algebra
\bea
i \left\{ L_n,L_m\right\} & = & (n-m)L_{n+m} + {c \over 12}(n^3-n)\delta_{n,-m} \, , \\
i \left\{ \overline{L}_n,\overline{L}_m\right\} & = & (n-m)\overline{L}_{n+m} +
 {c \over 12}(n^3-n)\delta_{n,-m} \, , \\
\left\{ L_n,\overline{L}_m\right\} & = & 0 \, ,
\eea
with central charge $c={3\ell \over 2G}$.

  Our aim now is to provide a more explicit description of the expansion (3-6). A 
simple and natural way to do this is to realize that the conformal symmetry
also arises from a conformal analysis of infinity. The conformally related metric 
$d\overline{s}^2 = {e^{2\phi} \over r^2}ds^2$ where $ds^2$
is the elementary black hole solution with $M=J=0$
\be
ds^2 = -r^2dx^+dx^- + {\ell^2 \over r^2}dr^2 \, ,
\ee 
and $e^{2\phi}$ is a positive function depending on the coordinates $x^{\pm}$,
induces a metric on the boundary $r\rightarrow\infty$
\be
d\overline{s}_b = -e^{2\phi}dx^+dx^- \, , \label{metric2}
\ee
The 2d conformal symmetry (7-8) acts naturally on the metric (\ref{metric2})
preserving the conformal structure. This is in fact a particular case of the
more general $AdS_{d+1}/CFT_d$ correspondence suggested in 
\cite{Maldacena} and elaborated in \cite{GKP,Witten2}. In Ref. \cite {HS} the Weyl
anomaly for conformal field theories described via the supergravity action 
in d=2,4,6 has been calculated regularising the gravitational action in a 
generally covariant way. In the case  d=2  one recovers the well-known trace anomaly
$\langle T^{\alpha}_{\alpha}\rangle = {c \over 24\pi}R$.
However, it was pointed out in \cite{Jackiw} that in quantising a $2d$ conformal field one can 
alternatively preserve the Weyl symmetry and partially break diffeomorphism invariance.
Therefore, the trace anomaly disappear but one encounters that the stress tensor does not obey the covariant conservation law:
\be
\langle\nabla_{\mu}T^{\mu\nu}\rangle = -{c \over 48\pi}\partial_{\nu}R(\sqrt{-g}g^{\alpha\beta}) \, .
\ee
In conformal coordinates this equation implies that $\langle T^{\mu\nu}\rangle$ 
transforms according to the Virasoro anomaly \cite{NNT}.
Our approach is related to this second viewpoint because, as we have already mentioned, the 
diffeomorphism generated by $L_n, n\neq0,\pm1$  should not be considered as 
"pure gauge transformations" due to the Virasoro anomaly. Therefore it is 
natural in our scheme to pick up a trivial metric on the boundary to represent the 
unique conformal equivalence class. However, due to the breakdown of general 
covariance in the boundary, we cannot choose a particular form of the flat metric,
 but instead we have to consider the general form of a flat metric. All this means that    
$\phi$ should be a free field.
  Going back to the bulk part of AdS$_3$ we see immediately that the metric
\be
ds^2 = -e^{2\phi}r^2 dx^+dx^- + {\ell^2 \over r^2}dr^2 \, , \label{metric3}
\ee 
induces the boundary metric (\ref{metric2}) and satisfy Einstein 's equations. 
However (\ref{metric3}) does not satisfy the asymptotic conditions (3-6). To recover 
these conditions we have to transform (\ref{metric3}) in an 
appropriated way. A redefinition of the radial coordinate $r\longrightarrow re^{-\phi}$
brings the metric into the form
\bea
ds^2 = &-&r^2dx^+dx^-+ {\ell^2 \over r^2}dr^2 - {2\ell^2 \over r}(\partial_+\phi dx^+dr +
\partial_-\phi dx^-dr) \nonumber \\
 &+&\ell^2(\partial_+\phi)^2(dx^+)^2 + \ell^2(\partial_-\phi)^2(dx^-)^2 \nonumber \\
&+& 2\ell^2(\partial_+\phi)(\partial_-\phi)dx^+dx^- \, , \label{step1}
\eea
The above metric fulfils the conditions (\ref{g1},\ref{g2},\ref{g4}), but to
satisfy the requirement (\ref{g3}) it is necessary
to do a second transformation $x^{\pm} \longrightarrow  x^{\pm} +
{\ell^2 \over r^2}\partial_{\mp}\phi$.
We then get
\bea
ds^2 = &-&r^2dx^+dx^-+ {\ell^2 \over r^2}dr^2 + (4{\ell^4 \over r^4}\partial_+\phi\partial_-\phi
+ {\cal O}({1 \over r^6}))dr^2 \nonumber \\
&+&(2\ell^2(\partial_+\phi\partial_-\phi-\partial_+\partial_-\phi) 
+ {\cal O}({1 \over r^2}))dx^+dx^- \nonumber \\
&+&(\ell^2((\partial_+\phi)^2-\partial_+^2\phi)
+ {\cal O}({1 \over r^2}))(dx^+)\nonumber \\
&+&(\ell^2((\partial_-\phi)^2-\partial_-^2\phi)
+ {\cal O}({1 \over r^2}))(dx^-)^2 \nonumber \\
&+&{\cal O}({1 \over r^3})dx^+dr + {\cal O}({1 \over r^3})dx^-dr \, , \label{step2}
\eea
where the omitted terms can be computed in a recursive way.
  Therefore we have
\bea
\gamma_{\pm\pm} & = & \ell^2((\partial_{\pm}\phi)^2-\partial_{\pm}^2\phi) \, , \\
\gamma_{+-} & = & \ell^2(\partial_+\phi\partial_-\phi-\partial_+\partial_-\phi) \, , \\
\gamma_{rr} & = &  4\ell^4(\partial_+\phi\partial_-\phi) \, , \\
%\gamma_0 & = & \gamma_{+-} - {1 \over 4\ell^2}\gamma_{rr} = -\ell^2(\partial_+\partial_-\phi) \, ,
\eea
but, as we have already mentioned, only the terms $\gamma_{\pm\pm}$, $\gamma_{+-} - {1 \over 4\ell^2}\gamma_{rr}$
are gauge invariant.

  The transformation law (\ref{delta}) can be reproduced if $\phi$ transforms as
\be
\delta\phi = \ell^2\left\{ 2\partial_+ \phi T^+ + 2\partial_- \phi T^- +
\partial_+ T^+ + \partial_- T^- \right\} \, ,
\ee
  Redefining now the scalar field $\overline{\phi}=\sqrt{\ell \over 2G}\phi$
the stress tensor $\Theta_{\pm\pm}$ takes the form
\be
\Theta_{\pm\pm} = {1 \over 2}\Big[ (\partial_{\pm}\overline{\phi})^2 -
\sqrt{\ell \over 2G}\partial_{\pm}^2 \overline{\phi} \Big] + {\ell \over 16G} \, ,
\label{qst}
\ee 
which is similar to that arising in Liouville theory. However, it is well-known \cite{BCT,HJ} that a
B\"acklund transformation convert the Liouville theory into an improved free field theory. It is just
this free field theory which emerges in this approach. 
%\be
%\partial_+\partial_-\overline{\phi} = 0 \, , \label{ffeq}
%\ee
The quantum stress tensor (\ref{qst}) gives rise to a central charge
\be
c = 1+3Q^2 \, , \label{c2}
\ee
where the background charge is given by
$Q =  \sqrt{\ell \over 2G} + 2\sqrt{2G \over \ell}$.                           
In a semiclassical description $\ell\gg G$ the central charge (\ref{c2})
reproduces the classical value (\ref{central}). It is also worthwhile
to remark that when $\phi$ is a chiral field there is not subleading
terms in (\ref{step2}) and one recovers solutions of the form (\ref{chiral}).

If we choose a boundary metric of constant curvature $\lambda$ to represent the given conformal equivalence class
\be
d\overline{s}_b = -e^{2\phi_L}dx^+dx^- \, , \label{metricL}
\ee
,where $\phi_L$ obeys a Liouville equation $\partial_-\partial_+{\phi_L} = {\lambda\over 8}e^{2\phi_L}$, the 
extension of the metric to the bulk part is more involved than (\ref{metric3}). Moreover, the first gauge invariant
subleading terms in the asymptotic expansion are 
\bea
\gamma_{\pm\pm} & = & \ell^2((\partial_{\pm}\phi_L)^2-\partial_{\pm}^2\phi_L) \, , \label{STL} \\ 
\gamma_{+-} - {1 \over 4\ell^2}\gamma_{rr} & = & -\ell^2(\partial_+\partial_-\phi_L - {\lambda\over 8}e^{2\phi_L}) \, ,
\eea
So we also recover the usual expression for the stress tensor of Liouville theory. Therefore,
and according to this, the freedom in picking a metric on the conformal structure seems to be related
to canonical transformations of the boundary theory. The boundary metric (\ref{metric3})
is the most natural one to relate the bulk gravity theory to the boundary conformal field theory.

  In this paper we have provided an explicit relation between the "would-be gauge" degrees of freedom
of the gravity theory and the underlying boundary conformal field theory. To do this we have made
use of the holographic correspondence AdS/CFT of \cite{Witten2} for the theory (\ref{action}).
However, 
instead of breaking Weyl invariance on the boundary as in \cite{Witten2,HS}, we sacrifice 
partially diffeomorphism invariance to make contact with the asymptotic boundary conditions of
\cite{BH,Strominger}. This way we have provided an explicit relation between the bulk metric on AdS$_3$,
verifying the boundary conditions of \cite{BH,Strominger} and an improved free field theory on the 
boundary. This result raises a question also pointed out by Carlip \cite {Carlip2} concerning the
derivation of the BTZ black hole entropy of Ref \cite{Strominger}. The Cardy's formula for the
asymptotic density of states of a conformal field theory with central charge $c$ 

\be
\log {\rho} (\Delta ,\overline{\Delta}) \sim 2\pi \sqrt{c \Delta \over 6} +
2\pi \sqrt{c \overline{\Delta} \over 6}
\ee
where $\Delta$ and $\overline{\Delta}$ are the eigenvalues 
of the two Virasoro generators $L_0 $ and $\overline{L}_0 $,
assumes that the lowest eigenvalues ($\Delta_0$,$\overline{\Delta}_0$)
of $L_0$ and $\overline{L}_0$ vanish.
In general, the above formula is still essentially valid if one replace the central charge "$c$"
by the so-called effective central charge $c_{eff}=c-24\Delta_0$ \cite{KS}.
However, the minimum value of $L_0$ ($\overline{L}_0$) is not zero for the improved free field 
$\overline{\phi}$,
in fact $c_{eff}=1$ as one should expect by a direct counting of states. Moreover, for normalizable
(macroscopic) states \cite{Seiberg} of Liouville theory we also have $c_{eff}=1$.   
To properly account for the black hole entropy one should include states of imaginary momentum
for the improved free field, or microscopic states in the terminology of Liouville theory.

In a recent paper \cite{GKS} it is claimed that stringy degrees of freedom  account for the
full density of states giving rise the Bekenstein-Hawking entropy of BTZ black holes.   

\section*{Acknowledgements}
P. Navarro acknowledges the Ministerio de Educaci\'on y Cultura for a FPU fellowship.

\end{document}